\documentclass[aps,prl,10pt,twocolumn,amsfonts,superscriptaddress,showpacs,floatfix]{revtex4-1}

\usepackage{color,verbatim}
\usepackage{bm}
\usepackage{bbold}
\usepackage{amsmath}
\usepackage{amssymb}
\usepackage{epsfig}
\usepackage{feynmf}
\usepackage{braket}
\usepackage{ulem}

\newcommand{\be}{\begin{equation}}
\newcommand{\ee}{\end{equation}}
\newcommand{\bea}{\begin{eqnarray}}
\newcommand{\eea}{\end{eqnarray}}
\newcommand{\bw}{\begin{widetext}}
\newcommand{\ew}{\end{widetext}}
\newcommand{\ba}{\begin{aligned}}
\newcommand{\ea}{\end{aligned}}

\newcommand{\dg}{^\dagger}

\renewcommand{\vec}[1]{\underline{\bm #1}}

\newcommand{\rd}{{\rm d}}

\def\nn{\nonumber\\}

\def\fr#1{(\ref{#1})}
\def\ocite#1{[\onlinecite{#1}]}

\def\g{\gamma}

\usepackage[colorlinks,bookmarks=false,citecolor=blue,linkcolor=blue,hyperfootnotes=true]{hyperref}

\begin{document}

\title{Prethermalization and thermalization in models with weak
  integrability breaking}

\author{Bruno Bertini}
\affiliation{The Rudolf Peierls Centre for Theoretical Physics, University of Oxford, Oxford, OX1 3NP, United Kingdom}
\author{Fabian H.L. Essler}
\affiliation{The Rudolf Peierls Centre for Theoretical Physics, University of Oxford, Oxford, OX1 3NP, United Kingdom}
\author{Stefan Groha}
\affiliation{The Rudolf Peierls Centre for Theoretical Physics, University of Oxford, Oxford, OX1 3NP, United Kingdom}
\author{Neil J. Robinson}
\affiliation{The Rudolf Peierls Centre for Theoretical Physics, University of Oxford, Oxford, OX1 3NP, United Kingdom}
\affiliation{Condensed Matter Physics and Materials Science Department, Brookhaven National Laboratory, Upton, New York 11973, USA}

\date{\today}

\begin{abstract}
We study the effects of integrability breaking perturbations on the
non-equilibrium evolution of many-particle quantum systems. We focus
on a class of spinless fermion models with weak interactions.
We employ equation of motion techniques that can be viewed as
generalizations of quantum Boltzmann equations. We benchmark our
method against time dependent density matrix renormalization group
computations and find it to be very accurate as long as interactions
are weak. For small integrability breaking, we observe robust
prethermalization plateaux for local observables on all accessible
time scales. Increasing the strength of the integrability breaking
term induces a ``drift'' away from the prethermalization plateaux
towards thermal behaviour. We identify a time scale
characterizing this cross-over. 
\end{abstract}

\maketitle
In classical mechanics, integrable few-particle systems can be
understood in terms of periodic, non-ergodic motion in action-angle
variables. Breaking integrability by adding a weak perturbation
induces a fascinating crossover between integrable and chaotic motion, 
which is described by the celebrated KAM theory \cite{KAM}. In
essence, classical few-particle systems with weak 
integrability breaking retain aspects of integrable motion on
intermediate time scales. Recently it has emerged, that similar
behaviour occurs in the non-equilibrium evolution of isolated
many-particle quantum systems. Starting with the seminal work of Rigol 
et al \cite{RigolPRL07} it has become clear that there is a dramatic
difference between the late time behaviour of isolated integrable and
non-integrable quantum many particle systems prepared in
initial states that are not eigenstates of the Hamiltonian. Generic
systems \textit{thermalize} \cite{RigolPRL07,DeutschPRA91,SrednickiPRE94,RigolPRL09,
RigolNature08,BiroliPRL10,RS10,BanulsPRL11,PolkonikovRMP11,aditi,rigol14,GE15}, i.e. exhibit relaxation of local
observables towards a Gibbs ensemble with an effective temperature,
while integrable systems evolve towards a generalized Gibbs ensemble
\cite{RigolPRL07,PolkonikovRMP11,GE15,RigolPRA06,CazalillaPRL06,CalabreseJStatMech07,
CramerPRL08,BarthelPRL08,FiorettoNJP10,CEF,FE,
EsslerPRL12,CauxPRL12,ColluraPRL13,MussardoPRL13,
PozsgayJStatMech13,FagottiJStatMech13,WoutersPRL14,PozsgayPRL14,
KormosPRA14,DeNardisPRA14,FagottiPRB14,SotiriadisJStatMech14,
GoldsteinArxiv14,EsslerArxiv14}. Starting with the work of Moeckel and
Kehrein~\cite{MoeKeh} it was then realized that models with weak
integrability breaking perturbations exhibit transient behaviour, in
which local observables relax towards non-thermal values that retain
information of the proximate integrable theory. This has been termed
\textit{prethermalization}, and has been established to occur in
several models \cite{MoeKeh,RoschPRL08,KollarPRB11,worm13,MarcuzziPRL13, EsslerPRB14,NIC14,Fagotti14,konik14,BF15,knap15}. Crucially, it was recently observed in experiments on
ultra-cold bosonic atoms \cite{Getal1,Getal2,Getal3}.  The general
expectation is that prethermalization is a transient effect, and at
``sufficiently late times'' non-integrable systems thermalize. While
this appears natural, there is scant evidence in support of this
scenario. The reason is that available numerical \cite{SWreview} 
or analytical \cite{{MoeKeh},{EsslerPRB14},BF15} methods are not able to
reach late enough times. The exception is the case of infinitely many
dimensions, where it was shown in a particular example that a weakly
non-integrable model thermalizes \cite{StarkArxiv13}. 
Here we address these issues in the context of
weakly interacting one dimensional many-particle systems. This case
has the important advantage that the accuracy of approximate methods
can be benchmarked by comparisons with powerful numerical methods like
the time dependent density matrix renormalization group (t-DMRG)
\cite{SWreview}. Moreover, the existence of many
strongly interacting one dimensional integrable systems makes it
possible to verify that the qualitative behaviour we find persists
for arbitrary interaction strengths. 

We focus on the weak interaction regime $U\alt J_1$ of the
three-parameter family of spinless fermion Hamiltonians  
\bea
H(J_2,\delta,U) &=& -J_1\sum_{l=1}^L \Big[1+(-1)^l\delta \Big] \Bigl( c\dg_l
c^{\phantom\dagger}_{l+1}+ {\rm H.c.}\Bigr)\nn 
&-&J_2\sum_{l=1}^L  \Bigl[ c\dg_l c^{\phantom\dagger}_{l+2}+ {\rm H.c.}
\Bigr]+{U}\sum_{l=1}^L n_ln_{l+1}.\ \
\label{Eq:Ham}
\eea
Here $c_i$ and $c^\dagger_i$ are spinless fermion operators on site
$i$ and the hopping amplitudes describe nearest-neighbor and
next-nearest-neighbor hopping respectively, while
$0\leq\delta<1$ is a dimerization 
parameter. Finally there is a repulsive nearest neighbour interaction
of strength $U$. From here onwards we set $J_1=1$ and measure all the
energies in units of $J_1$.  There are a several limits in which
  \fr{Eq:Ham} 
becomes integrable: (i) $U=0$ describes a free theory; (ii) $\delta =
J_2 = 0$ corresponds to the anisotropic spin-1/2 Heisenberg
chain~\cite{Orbach}; (iii) the
low-energy degrees of freedom for $J_2 = 0$ and $\delta,U \ll 1$ are
described by the quantum sine-Gordon model \cite{EKreview}. Away from
these limits, the model is non-integrable.   
Our protocol for inducing and analyzing non-equilibrium dynamics is as follows. We
prepare the system in an initial density matrix $\rho_0$ that is not an
eigenstate of $H(J_2,\delta,U)$ for any value of $U$. We then compare
the expectation values of local operators for time evolution with
the integrable $H(J_2,\delta,0)$ and (weakly) non-integrable $H(J_2,\delta,U)$
respectively. For $U=0$ our model is non-interacting,
and concomitantly in the thermodynamic limit expectation values of
local operators relax to time independent values described by a
generalized Gibbs ensemble. In the following we analyze how a small
integrability breaking interaction $U>0$ changes the non-equilibrium
evolution.  We stress that our protocol differs in a very important
way from the weak interaction quenches analyzed previously
\cite{{StarkArxiv13},{NessiArxiv15}}. In these works there is no
dynamics at all for $U=0$. Hence quenching the interaction from zero
to a finite value simultaneously breaks integrability and induces a
time dependence into the problem. This masks the interaction induced
modification of the integrable post-quench dynamics. Quantum
quenches in the model \fr{Eq:Ham} 
with $J_2=0$ were previously studied in Ref.~\ocite{EsslerPRB14} by
numerical and analytical methods. On the accessible time scales robust
prethermalization was observed, but no evidence for eventual
thermalization was found. While our manuscript was being completed a
paper appeared in which techniques similar to the ones we employ here
were used to analyze quantum quenches in the case $\delta_i=\delta_f = 0$
\cite{NessiArxiv15}. No prethermalization in our sense was observed for the
aforementioned reason that there is no dynamics without integrability
breaking in this case, but instead evolution towards a thermal steady
state was found. Given that $U$ is small, a convenient basis for
analyzing quench dynamics is obtained by diagonalizing the quadratic
part of the Hamiltonian. This results in 
\bea
H(J_2,\delta,0) &=& \sum_{\alpha=\pm}\sum_{k > 0} \epsilon_\alpha(k)
a\dg_\alpha(k) a_\alpha(k),
\eea
where $a_\pm(k)$ are momentum space annihilation operators
obeying canonical anticommutation relations
$\{a_\alpha(k),a^\dagger_\beta(q)\}=\delta_{\alpha,\beta}\delta_{k,q}$,
and $\epsilon_\alpha(k) = -2J_2\cos(2k) + 2\alpha
\sqrt{\delta^2+(1-\delta^2)\cos^2(k)}$ are single particle dispersion 
relations of the two bands of fermions. The system is initially (at
time $t=0$) prepared in a density matrix $\rho_0$, and subsequently
evolves according to
\be
\rho(t)=e^{-iH(J_2,\delta_f,U)t}\rho_0e^{iH(J_2,\delta_f,U)t}.
\ee
Using equation of motion (EOM) techniques \cite{{StarkArxiv13},{INJPhys}}
analogous to the ones employed in derivations of quantum Boltzmann
equations \cite{{ErdosJStatPhys04},{SpohnJStatPhys09}}, we obtain
evolution equations for the two-point functions  
\be
n_{\alpha\beta}(q,t) = \textrm{Tr}[\rho(t)a\dg_{\alpha}(q) a_{\beta}(q)]\,.
\label{Eq:Defnab}
\ee
The EOM can be cast in the form
\bw
\bea
\dot{n}_{\alpha\beta}(k,t) &=& i {\epsilon}_{\alpha\beta}(k)
n_{\alpha\beta}(k,t)+4iUe^{i  t{\epsilon}_{\alpha\beta}(k)} \sum_{\g_1} J_{\gamma_1\alpha}(k;t) n_{\g_1\beta}(k,0)-J_{\beta \g_1}(k;t) n_{\alpha\g_1}(k,0)\notag\\
&& - U^2  \int_0^t \rd t'  \sum_{\bm \gamma}\sum_{k_1,k_2{>0}} K^{\bm\g}_{\alpha\beta}(k_1,k_2; k; t-t') n_{\g_1\g_2}(k_1,t') n_{\g_3\g_4}(k_2,t')\nn
&& -U^2  \int_0^t \rd t'  \sum_{\vec{\g}}\sum_{k_1,k_2,k_3{>0}} L^{\vec{\g}}_{\alpha\beta}(k_1,k_2,k_3; k; t-t') 
n_{\g_1\g_2}(k_1,t') n_{\g_3\g_4}(k_2,t')n_{\g_5\g_6}(k_3,t'),
\label{Eq:EOM}
\eea
\ew
where $\epsilon_{\alpha\beta}(k)=\epsilon_{\alpha}(k)-\epsilon_{\beta}(k)$.
Explicit expressions for the
kernels  $J$, $K$, $L$ and details of our derivation are given in the
Supplemental Material. The solution of the set of integro-differential
equations~\fr{Eq:EOM} is numerically demanding. We designed an
algorithm that scales as $L^3\times T$ where $T$ is the number of time
steps and $L$ the number of lattice sites. This allows us to reach
long times $J_1t\sim 80$ on large systems $L\sim 320$ (a similar
scaling was proposed in Ref.~\ocite{NessiArxiv15}). Given the expectation values
\fr{Eq:Defnab}, we may readily calculate the single-particle Green's
function 
\bea
{\cal G}(j,l;t) &=&\textrm{Tr}[\rho(t)c\dg_j c^{\phantom{\dagger}}_l]\nn
&=&\frac{1}{L} \sum_{k>0} \sum_{\alpha,\beta = \pm} 
\gamma^*_\alpha(k,j) \gamma_\beta(k,l) n_{\alpha\beta}(k,t), 
\label{Eq:GreenFun}
\eea 
where the coefficients $\gamma_\alpha(k,j)$ are given in the
Supplementary Material. A crucial check of the accuracy of our
approach is provided by a direct comparison to previous t-DMRG
computations~\ocite{EsslerPRB14}.  
\begin{figure}[ht]
\includegraphics[width=0.48\textwidth]{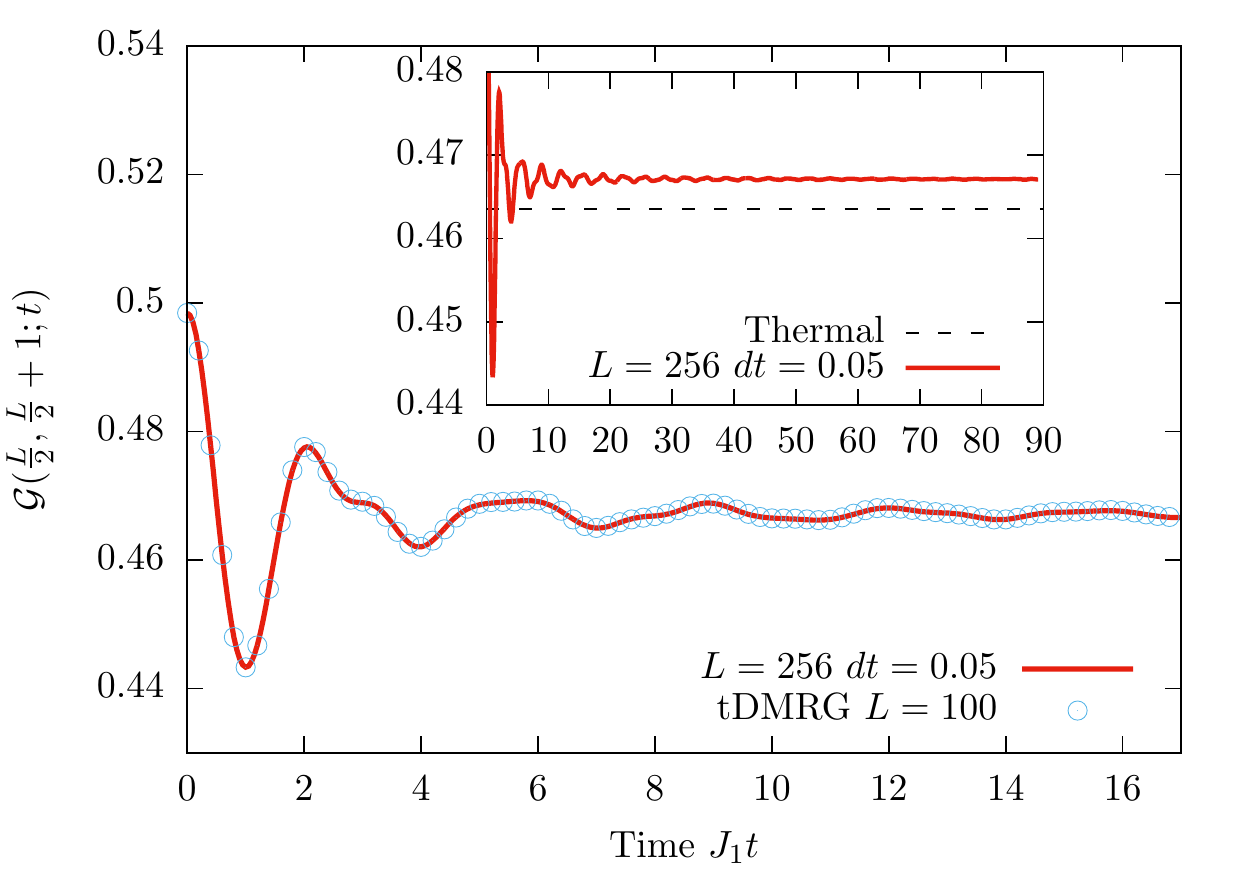}
\caption{(Color online) 
${\cal G}(\frac{L}{2},\frac{L}{2}+1;t)$ for a quench where the system
is prepared in the ground state of $H(0,0.8,0)$ and time evolved
with $H(0,0.4,0.4)$ for a system with $L=256$ sites. The EOM results
(red line) are in excellent agreement with t-DMRG computations \cite{EsslerPRB14}
(circles). Inset: prethermalized behaviour persists over a large time
interval.} 
\label{Fig:G1}
\end{figure} 
In Fig.~\ref{Fig:G1} we present a comparison of 
${\cal G}(L/2,L/2+1)$ between EOM and t-DMRG results for a quench
where the system is prepared in the ground state of $H(0,0.8,0)$ and
time evolved subject to the Hamiltonian $H(0,0.4,0.4)$. We see that
even for relatively large $U=0.4$, there is excellent agreement
between the two methods for all times accessible by t-DMRG. Similar
levels of agreement are found for other ${\cal G}(L/2,L/2+j)$ with
$j=2,3,4,5$. This agreement suggests that the EOM method is very
accurate for small values of $U$ and short and intermediate time
scales. The advantage of the EOM method is that it allows us to access
later time scales than the t-DMRG computations reported in
Ref.~\cite{EsslerPRB14}. As long as the interaction strength $U$ is
sufficiently small, we observe very long-lived prethermalization
plateaux, as is exemplified in the inset in Fig.~\ref{Fig:G1}. There,
the thermal value has been computed by quantum Monte Carlo simulations
on a system with $L=100$ sites.
\begin{figure}[ht]
\includegraphics[width=0.48\textwidth]{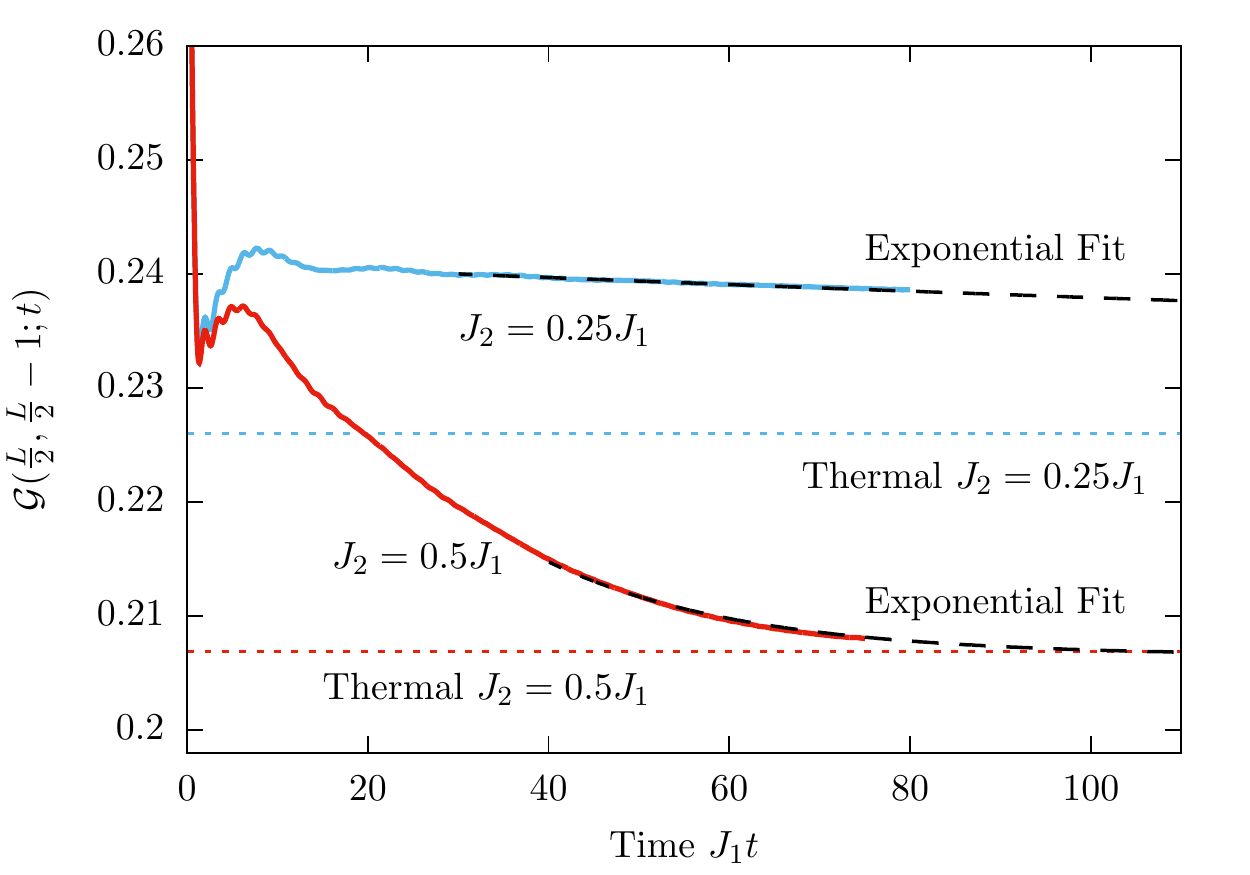}
\caption{(Color online) 
${\cal G}(\frac{L}{2},\frac{L}{2}-1;t)$ for a system with Hamiltonian
$H(J_2,0.1,0.4)$ and sizes $L=360, 320$ initially prepared in a thermal
state \fr{thermal} with density matrix $\rho(2,0,0,0)$.
The expected steady state thermal values
are indicated by dotted lines, while the black dashed lines are
exponential fits to \fr{Eq:ExpFit}.} 
\label{fig:TPG1}
\end{figure}

\begin{figure}[ht]
\includegraphics[width=0.48\textwidth]{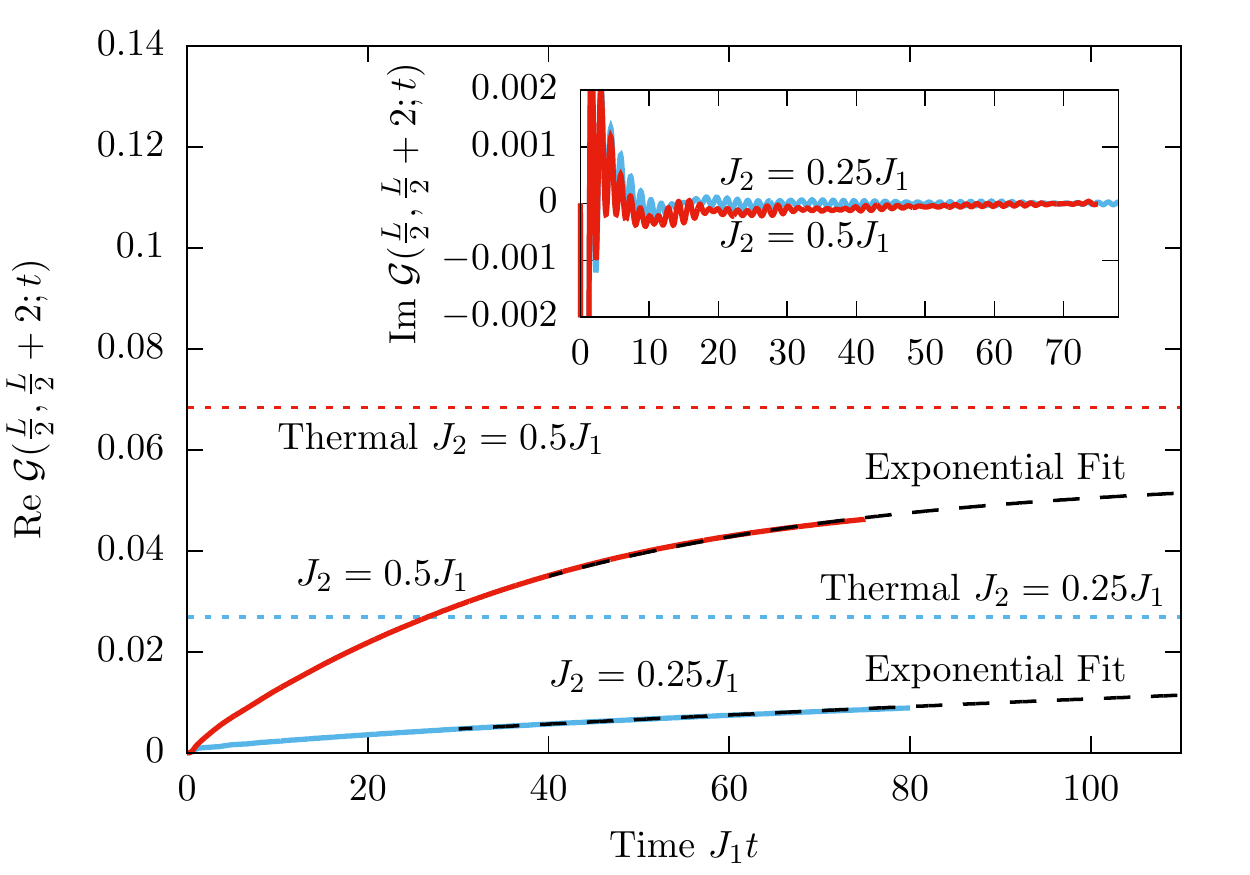}
\caption{(Color online) 
Real (Inset: imaginary) part of ${\cal
  G}(\frac{L}{2},\frac{L}{2}+2;t)$ 
for a system with Hamiltonian $H(J_2,0.1,0.4)$ and sizes $L=360, 320$,
that was initially prepared in a thermal state \fr{thermal} with
density matrix $\rho(2,0,0,0)$. The expected steady state thermal values
are indicated by dotted lines, while the black dashed lines are
exponential fits to \fr{Eq:ExpFit}. }
\label{Fig:TPG2}
\end{figure}

In order to investigate if and how the prethermalized regime
evolves towards thermal equilibrium it is convenient to invoke a
non-zero $J_2$. In essence, $J_2$ allows us to tune the
cross-over time scale between the two regimes. In order to access the
dynamics for a larger range of energy densities we consider thermal
initial density matrices of the form
\be
\rho( \beta,J_2,\delta,U)=\frac{e^{-\beta H(J_2,\delta,U)}}
{{\rm Tr}(e^{-\beta H(J_2,\delta,U)})}.
\label{thermal}
\ee
Figs.~\ref{fig:TPG1} and \ref{Fig:TPG2} show results for the time evolution
of the Green's function for a system prepared in the initial state
\fr{thermal} with density matrix $\rho(2,0,0,0)$, and time evolved
with Hamiltonian $H(J_2,0.1,0.4)$.  In contrast to the case $J_2=0$,
$U=0.4$, we now observe a slow drift towards a thermal steady
state. Increasing $J_2$ enhances the drift. The thermal values shown
in Figs.~\ref{fig:TPG1} and  \ref{Fig:TPG2} are obtained as follows. The
energy density is given by $e={\rm Tr}[\rho(2,0,0,0)H(J_2,0.1,0.4)]/L$ and determines the
effective temperature $1/\beta_{\rm eff}$ of the thermal ensemble
for the post-quench Hamiltonian $H(J_2,0.1,0.4)$ through $e={\rm
  Tr}[\rho(\beta_{\rm eff},J_2,0.1,0.4)H(J_2,0.1,0.4)]/L$ \cite{note:trace}. We determine $\beta_{\rm
eff}$ by exact diagonalization of small systems up to size $L=16$, and
then use the same method to compute the single-particle Green's
function in thermal equilibrium at temperature $1/\beta_{\rm eff}$. We
note that ${\cal G}(i,j;t) $ is real for odd separations
$|i-j|$. For even $|i-j|$ the imaginary part is non-zero but small and
relaxes towards zero. We find that the observed relaxation towards
thermal values is compatible with exponential decay
\be
{\cal G}(i,j;t) \sim {\cal G}(i,j)_{\text{th}} + A_{ij}(
J_2, \delta, U) e^{-t/\tau_{ij}(J_2, \delta, U)}\,, 
\label{Eq:ExpFit}
\ee
where ${\cal G}(i,j)_{\text{th}}$ is the thermal Green's function at
temperature $1/\beta_{\rm eff}$ \cite{note:imag}. The decay times $\tau_{ij}(J_2,
\delta, U)$ are quite sensitive to the value of $J_2$. This can be
understood by noting that large values of $J_2$ modify the band
structure of the non-interacting model by introducing additional
crossings at a fixed energy. This, in turn, generates additional
scattering channels that promote relaxation. 

\begin{figure}
\begin{tabular}{cc}
\includegraphics[width=0.45\textwidth]{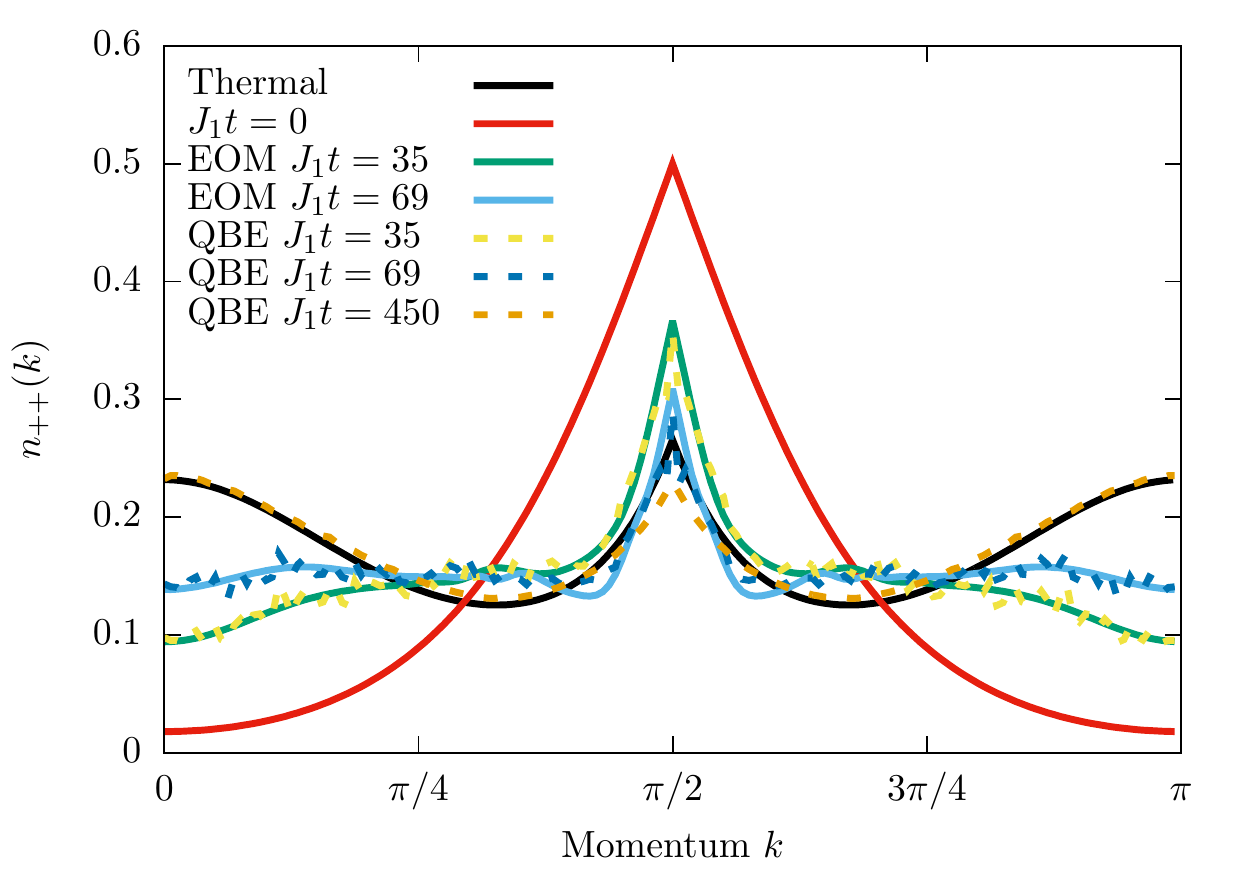} \\
\includegraphics[width=0.45\textwidth]{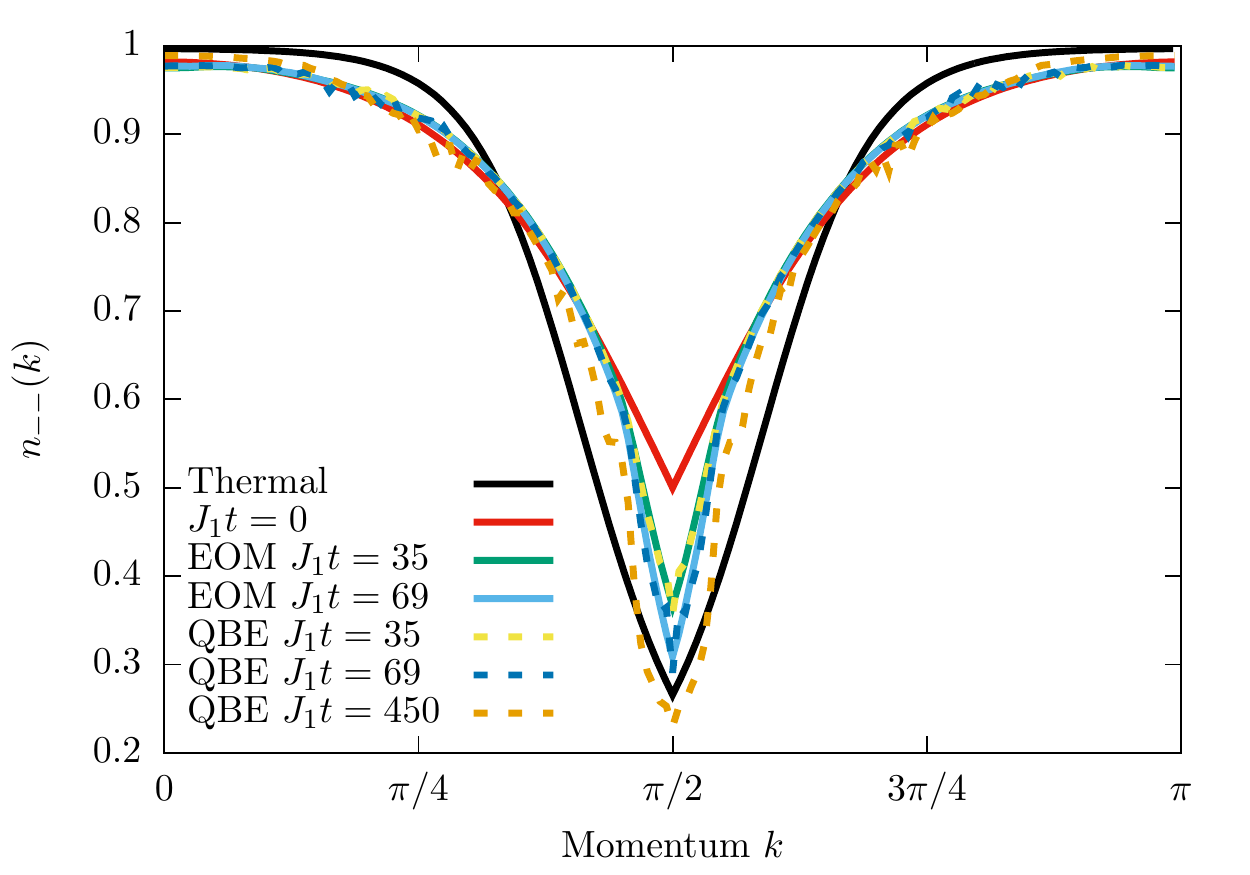}
\end{tabular}
\caption{(Color online) Occupation numbers $n_{++}(k,t)$ and $n_{--}(k,t)$
initialized in the thermal state \fr{thermal} $\rho(2,0,0.5,0)$, and
time evolved with $H(0.5,0,0.4)$. The solid lines are the results of
the EOM $(L=320)$ for various times. The dotted lines are computed by
means of the QBE
($L=320$). The black solid line is the thermal value found by means of second order perturbation theory in $U$.} 
\label{Fig:nks}
\end{figure}

A natural question is whether the integral equation
\fr{Eq:EOM} can be simplified in the late time regime by removing the
time integration, in analogy with standard quantum Boltzmann equations
(QBE) \cite{{ErdosJStatPhys04},{SpohnJStatPhys09}}. Here we are faced with
the difficulty that the structure of our EOM \fr{Eq:EOM} is quite
different from the ones studied in
Refs~\cite{{ErdosJStatPhys04},{SpohnJStatPhys09}}. However, in the  
case $\delta_f=0$ numerical integration of the full EOM \fr{Eq:EOM}
suggests that the ``off-diagonal'' occupation numbers become
negligible at late times $n_{+-}(k,t)\approx 0$ and it is possible to
derive a QBE for ``diagonal'' occupation numbers. The QBE for
$\delta_f=0$ reads
\bea
&&  \hspace{-0.15cm} \dot{n}_{\alpha\alpha}(k,\tau) =  - \sum_{\gamma,\delta}\sum_{p,q{>0}} \widetilde K^{\g\delta}_{\alpha\alpha}(p,q;k) n_{\g\g}(p,\tau) n_{\delta\delta}(q,\tau)\nn
&& \hspace{-0.3cm}-\sum_{\g,\delta,\epsilon}\sum_{p,q,r{>0}} \widetilde L^{\g\delta\epsilon}_{\alpha\alpha}(p,q,r;k)n_{\g\g}(p,\tau) n_{\delta\delta}(q,\tau)n_{\epsilon\epsilon}(r,\tau).
\label{Eq:BB}
\eea
Here $\tau = U^2 t$ is the usual rescaled time variable, $t_0\gg 1/U$
is the time at which the kinetic equation is initialized and the
functions $\tilde K$, $\tilde L$ are given in the Supplemental
Material.  The QBE agrees with the EOM for
sufficiently late times (an example is shown in Fig.~\ref{Fig:nks},
see the discussion below). Because of its simpler structure, the QBE allows us to
access later times than we able to reach with the EOM approach. In
particular, employing the QBE we conclude that for weak interactions
the relaxation times in \fr{Eq:ExpFit} scale as \cite{BEGRinprep}
\be
\tau^{-1}_{ij}(J_2, \delta_f=0, U)\propto U^2.
\ee
This is in contrast to the $U^4$ scaling found for interaction quenches in
the infinite dimensional Hubbard model \cite{StarkArxiv13}.

To establish more comprehensively that the integrability breaking
perturbation leads to thermalization, we consider the (Bogoliubov) mode
occupation numbers $n_{\alpha\beta}(q,t)$ themselves. The mode
occupation operators are not local in space, and hence it is not a
priori clear that their expectation values should eventually
thermalize; see however Ref.~\cite{wright14}. Importantly, we only
consider initial states with finite correlation lengths, which implies
that ${\cal G}(j,l;t)$ are exponentially small in $|j-l|$ as long as
$|j-l|\gg J_1 t$ \cite{Bravyi}. This, together with the fact that
${\cal G}(j,l;t)$ decay exponentially fast in time for $|j-l|\leq J_1
t$, suggests that $n_{\alpha\beta}(q,t)$ should relax in the regime
$1\ll J_1t \ll L$.  
In Fig.~\ref{Fig:nks} we present the mode occupation numbers
$n_{\alpha\alpha}(k,t)$ at several different times for a system of
size $L=320$ prepared in the density matrix $\rho(2,0,0.5,0)$ and
evolved with Hamiltonian $H(0.5,0,0.4)$. For short and intermediate
times $J_1t<70$ we use the full EOM, while late times are accessible
only to the QBE. The QBE is initialized at time $t_0=20$, and is seen
to be in good agreement with the full EOM until the latest times
accessible by the latter method. We observe that at intermediate times
both $n_{++}(k,t)$ and $n_{--}(k,t)$ slowly approach their respective 
thermal distributions at the effective temperature $1/\beta_{\rm eff}$
introduced above. The ``off-diagonal'' occupation numbers
$n_{+-}(k,t)$, calculated by integrating the full EOM, approach their
thermal value zero in an oscillatory fashion. The observed behaviour
of the mode occupation numbers strongly suggests that the weak
integrability breaking term indeed induces thermalization. 

We note that in the QBE framework the final relaxation is towards the
non-interacting Fermi-Dirac distribution with an effective temperature
set by the kinetic energy at the time the Boltzmann is initialized
\cite{{SpohnJStatPhys09}, {SpohnPRE12}}, signalling the   
importance of corrections to the QBE at very late times. 
Such corrections, arising from higher cumulants, are important for
obtaining the power law behaviour expected at very late times (for
certain observables) after quenches in non-integrable models
\cite{Luxetal,kim}.

In this work we have developed a method that allows us to analyze the
effects of a weak integrability breaking interaction on the time
evolution of local observables after a quantum quench. We have shown
that there is a crossover between a prethermalized regime,
characterized by the proximity of our model to an integrable theory,
and a thermal steady state. The observed drift of ${\cal G}(i,j;t)$ in
time towards its thermal value is exponential and characterized by a
time scale proportional to $U^{-2}$. The models considered here feature a
global $U(1)$ symmetry (particle number conservation). A preliminary
analysis suggests that the scenario found here, a prethermalized
regime followed by a cross-over to a thermal steady state, occurs also
in absence of this $U(1)$ symmetry \cite{BEGRinprep}.

\acknowledgments
We thank M. Fagotti, A. Gambassi, S. Kehrein and A. Silva for helpful
discussions. This work was supported by the EPSRC under grants 
EP/J014885/1 and EP/I032487/1, by the U.S. Department of Energy,
 Office of Basic Energy Sciences, under Contract Nos. DEAC02-98CH10886
  and DE-SC0012704 (N.J.R.) and by the Clarendon Scholarship fund (S.G.).



\newpage ${}$ \newpage
\setcounter{equation}{0}
\setcounter{figure}{0}
\setcounter{table}{0}
\makeatletter
\renewcommand{\theequation}{S\arabic{equation}}
\renewcommand{\thefigure}{S\arabic{figure}}

\onecolumngrid

\begin{center}
{\large{\bf Supplemental Material for ``Prethermalization and thermalization in models with weak
  integrability breaking''}}
\end{center}

\section{Diagonalizing the non-interacting Hamiltonian}
\label{App:Bogo}
The non-interacting part of the Hamiltonian~\fr{Eq:Ham}
\bea
H_0(J_2,\delta) = - J_1 \sum_l [1+\delta(-1)^l](c\dg_l c^{\phantom\dagger}_{l+1} + c\dg_{l+1} c^{\phantom\dagger}_l) - J_2 \sum_l (c\dg_l c^{\phantom\dagger}_{l+2} + c\dg_{l+2} c^{\phantom\dagger}_l),\nonumber
\eea
is diagonalized by the canonical transformation
\be
c_l=\frac{1}{\sqrt{L}}\sum_{k>0}\sum_{\alpha=\pm} \gamma_\alpha(l,k|\delta)a_\alpha(k)\ ,
\label{Eq:BogoTrans}
\ee
where the coefficients are given by 
\be
\gamma_{\pm}(2j-1,k|\delta)=e^{-i k (2j-1)}\,,\qquad \gamma_{\pm}(2j,k|\delta)=\pm e^{-i k 2 j} e^{-i \varphi_k(\delta)}\,,\qquad
e^{-i \varphi_k(\delta)}=\frac{-\cos k+ i \delta\sin k}{\sqrt{\cos^2 k+ \delta^2\sin^2k}}\,.\nonumber
\ee
In the new basis we have
\bea
H_0(J_2,\delta) &=& \sum_{\alpha=\pm}\sum_{k > 0} \epsilon_\alpha(k)
a\dg_\alpha(k) a_\alpha(k)\ ,
\eea
where the single-particle dispersions are
\be
\epsilon_\alpha(k) = -2J_2\cos(2k) + 2\alpha J_1
\sqrt{\delta^2+(1-\delta^2)\cos^2(k)}.
\label{dispersionrels}
\ee

Applying the same transformation to the interaction part of the Hamiltonian
$H_{\rm int} = U\sum_{l=1}^L c\dg_l c_l c\dg_{l+1} c_{l+1}$ gives 
\be
H_{\rm int} = U \sum_{\bm\alpha}\sum_{\bm k {>0}} V_{\bm \alpha}({\bm k}) a\dg_{\alpha_1}(k_1)a\dg_{\alpha_2}(k_2)a_{\alpha_3}(k_3)a_{\alpha_4}(k_4).
\ee
Here we have introduced the notations ${\bm \alpha} =
(\alpha_1,\alpha_2,\alpha_3,\alpha_4)$, ${\bm k} = (k_1,k_2,k_3,k_4)$ and ${\bm k}>0$ is a shorthand notation for $k_i > 0$ for all $i=1,\ldots,4$. The interaction vertex factor can be written in a conveniently
antisymmetrized form
\bea
V_{\bm\alpha}({\bm k}) &=&- \frac{1}{4} \sum_{P,Q\in S_2} {\rm sgn}(P) {\rm sgn}(Q) 
V'_{\alpha_{P_1}\alpha_{Q_1}\alpha_{P_2}\alpha_{Q_2}}(k_{P_1},k_{Q_1},k_{P_2},k_{Q_2})\ ,\nn
V'_{\bm \alpha}(\bm k)&=&\frac{e^{i(k_3-k_4)}}{2L}\left(\alpha_1\alpha_2e^{i \varphi_{k_1}(\delta)}e^{-i \varphi_{k_2}(\delta)}+\alpha_3\alpha_4e^{i \varphi_{k_3}(\delta)}e^{-i \varphi_{k_4}(\delta)}\right)\delta_{k_1-k_2+k_3-k_4,0}\notag\\
&&\quad+\frac{e^{i(k_3-k_4)}}{2L}\left(\alpha_1\alpha_2e^{i
  \varphi_{k_1}(\delta)}e^{-i
  \varphi_{k_2}(\delta)}-\alpha_3\alpha_4e^{i
  \varphi_{k_3}(\delta)}e^{-i
  \varphi_{k_4}(\delta)}\right)\delta_{k_1-k_2+k_3-k_4\pm\pi,0}\ ,
\label{Int}
\eea
where $P=(P_1,P_2)$ and $Q=(Q_1,Q_2)$ are permutations of $(1,2)$ and
$(3,4)$ respectively.

\section{Equations of motion}
\label{App:EOM}
The equations of motion~\fr{Eq:EOM} are derived by following the steps
set out in Ref.~\cite{ErdosJStatPhys04} for deriving quantum
Boltzmann equations. The starting point are the Heisenberg equations of 
motion (EOM) for the fermion bilinears
$\hat{n}_{\alpha\beta}(q,t)=a^\dagger_\alpha(q,t)a_\beta(q,t)$ . They
are of the form
\begin{align}
\frac{\partial}{\partial
  t}\hat{n}_{\alpha\beta}(k,t)&=i\left[H,\hat{n}_{\alpha\beta}(k,t)\right]
=
i\left[{\epsilon}_\alpha(k,\delta)-{\epsilon}_\beta(k,\delta)\right]\hat{n}_{\alpha\beta}(k,t)+i
U\sum_{{\bm \alpha}}\sum_{{\bm{
    q}{>0}}}{Y}_{\alpha\beta}^{\boldsymbol{\alpha}}(k,\bm{q})\hat{A}_{\boldsymbol{\alpha}}(\bm{q},t)\ ,
\label{Heisenberg}
\end{align}
where we have defined
$\hat{A}_{\boldsymbol{\alpha}}(\bm{q},t)=a^{\dag}_{\alpha_1}(q_1,t)a^\dag_{\alpha_2}(q_2,t)a_{\alpha_3}(q_3,t)a_{\alpha_4}(q_4,t)$, and  
\be
{Y}_{\alpha\beta}^{\boldsymbol{\alpha}}(k,\bm{q})= \delta_{\beta,\alpha_4}\delta_{k,q_4}{V}_{\alpha_1\alpha_2\alpha_3\alpha}(\bm{q})+\delta_{\beta,\alpha_3}\delta_{k,q_3}{V}_{\alpha_1\alpha_2\alpha\alpha_4}(\bm{q})-\delta_{\alpha,\alpha_2}\delta_{k,q_2}{V}_{\alpha_1\beta\alpha_3\alpha_4}(\bm{q})-\delta_{\alpha,\alpha_1}\delta_{k,q_1}{V}_{\beta\alpha_2\alpha_3\alpha_4}(\bm{q})\,.\nonumber
\ee
In the second step we consider the Heisenberg equation of motion for
the operator $\hat{A}_{\boldsymbol{\alpha}}(\bm{q},t)$
\be
\frac{\partial}{\partial t}\hat{A}_{\boldsymbol{\alpha}}(\bm{q},t)=i\left[H,\hat{A}_{\boldsymbol{\alpha}}(\bm{q},t)\right]
=i {E}_{\boldsymbol{\alpha}}(\bm{q})\hat{A}_{\boldsymbol{\alpha}}(\bm{q},t)+i U\sum_{\boldsymbol{\gamma}}\sum_{\bm{p}{>0}} {V}_{\boldsymbol{\gamma}}(\bm{p})\left[\hat{A}_{\boldsymbol{\gamma}}(\bm{p},t),\hat{A}_{\boldsymbol{\alpha}}(\bm{q},t)\right]\,,
\label{Heisenberg4point}
\ee
where $ {E}_{\boldsymbol{\alpha}}(\bm{q})\equiv{\epsilon}_{\alpha_1}(q_1)+{\epsilon}_{\alpha_2}(q_2)-{\epsilon}_{\alpha_3}(q_3)-{\epsilon}_{\alpha_4}(q_4)$.
Integrating \fr{Heisenberg4point} in time and then taking an
expectation value with respect to our initial density matrix $\rho_0$, we have
\be
\braket{\hat{A}_{\boldsymbol{\alpha}}(\bm{q},t)}=\braket{\hat{A}_{\boldsymbol{\alpha}}(\bm{q},0)}e^{i  t {E}_{\boldsymbol{\alpha}}(\bm{q})}+i U\int_{0}^{t}ds \sum_{\boldsymbol{\gamma}}\sum_{\bm{p}{>0}} e^{i  (t-s){E}_{\boldsymbol{\alpha}}(\bm{q})}{V}_{\boldsymbol{\gamma}}(\bm{p})\braket{\left[\hat{A}_{\boldsymbol{\gamma}}(\bm{p},s),\hat{A}_{\boldsymbol{\alpha}}(\bm{q},s)\right]}\,. \nonumber
\ee
Substituting this back into \fr{Heisenberg} leads to an exact
integro-differential equation for the mode occupation numbers
$n_{\alpha\beta}(k,t)={\rm Tr}[\rho_0\hat{n}_{\alpha\beta}(k,t)]$,
which takes the form
\begin{align}
\dot{n}_{\alpha\beta}(k,t)=&i\left[{\epsilon}_\alpha(k,\delta)-{\epsilon}_\beta(k,\delta)\right] n_{\alpha\beta}(k,t)+i U\sum_{{\bm \alpha}}\sum_{{\bm{
    q}{>0}}}{Y}_{\alpha\beta}^{\boldsymbol{\alpha}}(k,\bm{q})\braket{\hat{A}_{\boldsymbol{\alpha}}(\bm{q},0)} e^{i  t {E}_{\boldsymbol{\alpha}}(\bm{q})} \notag\\
&-U^2\int_{0}^{t}ds\sum_{\boldsymbol{\alpha},\boldsymbol{\gamma}}\sum_{\bm{q},\bm{p}{>0}}\braket{\hat{A}_{\boldsymbol{\gamma}}(\bm{p},s)\hat{A}_{\boldsymbol{\alpha}}(\bm{q},s)} \left[{Y}_{\alpha\beta}^{\boldsymbol{\alpha}}(k,\bm{q}) e^{i  (t-s){E}_{\boldsymbol{\alpha}}(\bm{q})}{V}_{\boldsymbol{\gamma}}(\bm{p})-(\boldsymbol{\alpha},\bm{q})\rightarrow(\boldsymbol{\gamma},\bm{p})\right]\,.\label{EOM}
\end{align}
As Wick's theorem holds for all initial density matrices $\rho_0$ we
consider, the expectation value
$\braket{\hat{A}_{\boldsymbol{\alpha}}(\bm{q},0)}$ can be
expressed in terms of the mode occupation numbers
$n_{\alpha\beta}(k,0)$. The eight-point average in \fr{EOM} can be
decomposed as
\be
\braket{\hat{A}_{\boldsymbol{\gamma}}(\bm{p},t)\hat{A}_{\boldsymbol{\alpha}}(\bm{q},t)}=f(\{n_{\alpha\beta}(k,t)\})
+\mathcal C[\braket{\hat{A}_{\boldsymbol{\gamma}}(\bm{p},t)\hat{A}_{\boldsymbol{\alpha}}(\bm{q},t)}]\,, \nonumber
\ee
where the first term is the result of applying Wick's theorem,
and $\mathcal C\left[\cdots\right]$ denotes terms involving four, six
and eight particle cumulants (the eight particle cumulant does not
contribute because of the antisymmetric structure of \fr{EOM}). In
order to turn \fr{EOM}  into a closed system of 
integro-differential equations we now assume that the four and six
particle cumulants can be neglected at all times. This leads to the
following system of equations
\bea
\dot{n}_{\alpha\beta}(k,t) &=&i{\epsilon}_{\alpha\beta}(k) n_{\alpha\beta}(k,t)+4iUe^{i  t {\epsilon}_{\alpha\beta}(k)} \sum_{\g_1} J_{\gamma_1\alpha}(k;t) n_{\g_1\beta}(k,0)-J_{\beta \g_1}(k;t) n_{\alpha\g_1}(k,0)\notag\\
&& - U^2  \int_0^t \rd t'  \sum_{\bm \gamma}\sum_{k_1,k_2{>0}} K^{\bm\g}_{\alpha\beta}(k_1,k_2;k;t-t') n_{\g_1\g_2}(k_1,t') n_{\g_3\g_4}(k_2,t')\nn
&& -U^2  \int_0^t \rd t'  \sum_{\vec\g}\sum_{k_1,k_2,k_3{>0}} L^{\vec\g}_{\alpha\beta}(k_1,k_2,k_3;k;t-t')n_{\g_1\g_2}(k_1,t') n_{\g_3\g_4}(k_2,t')n_{\g_5\g_6}(k_3,t'),
\label{AB:EOM}
\eea
where $\vec\g=(\g_1,\ldots,\g_6)$ and we introduced the functions 
\bea
J_{\alpha\beta}(k;t)&=& e^{i \epsilon_{\alpha\beta}(k) t}\sum_{\g_2\g_3}\sum_{q{>0}} V_{\alpha \gamma_2 \gamma_3 \beta}(k,q,q,k) e^{i \epsilon_{\g_2\g_3}(q) t} n_{\g_2\g_3}(q,0)\,,\notag\\
K^{\bm\g}_{\alpha\beta}(k_1,k_2;k;t) &=& 4 \sum_{k_3,k_4{>0}} \sum_{\nu,\nu'} 
X^{\g_1\g_3\nu\nu';\nu\nu'\g_4\g_2}_{{\bm k};{\bm k}'} (\alpha,\beta;k;t),\nn
L^{\vec\g}_{\alpha\beta}(k_1,k_2,k_3;k;t) &=& 
8 \sum_{\nu}\sum_{k_4{>0}} X^{\g_1\g_3\g_6\nu;\nu\g_5\g_4\g_2}_{{\bm k};{\bm k}'}(\alpha,\beta;k;t)- 16\sum_{\nu} X^{\g_1\g_3\nu\g_4;\g_5\nu\g_6\g_2}_{k_1k_2k_1k_2;k_3k_1k_3k_1}(\alpha,\beta;k;t)\,,\nn
X^{{\bm\g};{\bm\alpha}}_{{\bm k};{\bm q}}(\alpha,\beta;q;t) &=&
Y^{\bm\g}_{\alpha\beta}({\bm k}|q)V_{\bm\alpha}({\bm q}) e^{i
  E_{\bm\g}({\bm k})t} - ({\bm \gamma},{\bm k})\leftrightarrow({\bm
  \alpha},{\bm q}).
\eea
The occupation numbers at time $t=0$ for a system prepared in an initial
state with density matrix $\rho(\beta,0,\delta_i,0)$ and time evolved
with Hamiltonian $H(J_2,\delta_f,U)$ are readily calculated using
Wick's theorem
\begin{align}
& n_{\alpha\alpha}(k)=\frac{1}{2}-\frac{1}{2}\cos(\varphi_k(\delta_f)-\varphi_k(\delta_i))\tanh( \beta\epsilon_\alpha^{(0)}(k)/2)\,,&\alpha=\pm\,,\\
&n_{\alpha\beta}(k)=\frac{
    i}{2}\sin({\varphi_k(\delta_f)}-\varphi_k(\delta_i))\tanh(
  \beta\epsilon_\alpha^{(0)}(k)/2)\,,&\alpha\neq\beta\ .
\label{AB:Occval}
\end{align}
Here the dispersions $\epsilon_\alpha^{(0)}(k)$ are given by
\fr{dispersionrels} with $J_2=0$ and $\delta=\delta_i$.

\section{Quantum Boltzmann equation (QBE) for $\delta_f=0$}
\label{App:Bolt}
The equations of motion obtained in our case are generally quite
different from known cases, in which (matrix) QBEs can be derived~\cite{ErdosJStatPhys04, SpohnJStatPhys09, SpohnPRE12}. 
An exception is the case $\delta_f=0$, where it is possible to obtain
a QBE for the ``diagonal'' occupation numbers $n_{\alpha\alpha}(q,t)$
as we will now show. Numerical integration of the full EOM, suggests that for small
interactions $U\ll1$ and at sufficiently late times $t>t_0 \sim
U^{-1}$ the occupation numbers $n_{\alpha\alpha}(k,t)$ depend only on
the variable $\tau \equiv U^2 t$, while $n_{+-}(k,t)\approx 0$. To
describe this late time regime it is convenient take the formal
scaling limit $U\rightarrow0$, $t\rightarrow\infty$ keeping
$\tau = t U^2$ fixed of the EOM~\fr{Eq:EOM}. In this limit the EOM
take the form
\bea
\dot{n}_{\alpha\alpha}(k,\tau) &=&\lim_{U\rightarrow0} 4iU^{-1} \sum_{\g_1} \left\{J_{\gamma_1\alpha}(k;\tau U^{-2}) n_{\g_1\alpha}(k,0)-J_{\alpha \g_1}(k;\tau U^{-2}) n_{\alpha\g_1}(k,0)\right\}\notag\\
&& +\lim_{U\rightarrow0} \sum_{\bm k{>0}} 
\int_0^{\tau} \frac{\rd \sigma}{U^2}\, e^{i E(\bm k)
  (\tau-\sigma)U^{-2}} F(\bm k;k; \sigma )\ , 
\label{AC:EOM}
\eea 
where we have collected most of the integrand of the $\sigma$-integral
into a single function $F(\bm k;k; \sigma)$ in order to lighten notations. 
The second contribution on the right hand side can be simplified by
using our assumptions that $n_{+-}(k,\sigma)\approx 0$ 
and $n_{\alpha\alpha}(k,\sigma)$ are slowly varying functions of $\sigma$
for $\sigma\agt U$. We thus have
\be
\int_0^{\tau} \frac{\rd \sigma}{U^2}\, e^{i E(\bm k)
  {(\tau-\sigma)}{U^{-2}}} F(\bm k;k; {\sigma})\approx\int_0^{U^{-1}} {\rd s}\, e^{i E(\bm k) (t-s)} F(\bm k;k; s)+ F(\bm k;k; \tau) \int_{U^{-1}}^{t}  \rd s\,  e^{i E(\bm k) (t-s)}\,.
\label{AD:simpint}
\ee
The first term vanishes in our scaling limit. We regularize the integral
in the second term by replacing $E(\bm k)\rightarrow E(\bm k)+i\eta$,
$\eta$ is small and positive
\bea
\lim_{U\to 0}\int_{U^{-1}}^{t} \rd s\, e^{i [E(\bm k)+i\eta] (t-s)}
=\frac{i}{E(\bm k)+i\eta}\equiv D( E({\bm k}))\,.
\label{AD:Reg}
\eea
The first contribution on the right hand side depends only on the
initial mode occupation numbers $n_{\alpha\beta}(k,0)$. In the case
$\delta_f=0$, the leading contribution at late times is obtained by
evaluating the momentum sums by a saddle point approximation. This gives
\be
\frac{4i}{U}\sum_{\g_1} \left\{J_{\g_1\alpha}(k;t) n_{\g_1\alpha}(k,0)-J_{\alpha \g_1}(k;t)
n_{\alpha\g_1}(k,0)\right\}\approx
\frac{\sin(\epsilon_{+-}(0) t)}{U
  t^{3/2}}(A_{\alpha}(k)e^{i\epsilon_{+-}(k) t}+\textrm{c.c.})\,,  
\label{AC:decay}
\ee
where $A_{\alpha}(k)$ is an amplitude depending on the initial
state and the vertex function. The right hand side of \fr{AC:decay} vanishes in the scaling
limit. Putting everything together, we obtain the following QBE in the
scaling limit
\bea
\dot{n}_{\alpha\alpha}(k,\tau) &=&-   \sum_{\bm \gamma}\sum_{k_1,k_2{>0}} \widetilde K^{\g_1\g_2}_{\alpha\alpha}(k_1,k_2;k) n_{\g_1\g_1}(k_1,\tau) n_{\g_2\g_2}(k_2,\tau)\nn
&& -   \sum_{\vec\g}\sum_{k_1,k_2,k_3{>0}} \widetilde L^{\g_1\g_2\g_3}_{\alpha\alpha}(k_1,k_2,k_3;k) 
n_{\g_1\g_1}(k_1,\tau) n_{\g_2\g_2}(k_2,\tau)n_{\g_3\g_3}(k_3,\tau)\,.
\label{ADEq:BB}
\eea
Here the kernels are given by
\bea
 \widetilde K^{\g_1\g_2}_{\alpha\beta}(k_1,k_2|q) &=& 4 \sum_{k_3,k_4{>0}} \sum_{\nu,\nu'}  \widetilde X^{\g_1\g_2\nu\nu'|\nu\nu'\g_2\g_1}_{{\bm k}|{\bm k}'} (\alpha,\beta|q),\nn
\widetilde L^{\g_1\g_2\g_3}_{\alpha\beta}(k_1,k_2,k_3|q) &=& 8 \sum_{\nu}\sum_{k_4{>0}} \widetilde X^{\g_1\g_2\g_3\nu|\nu\g_3\g_2\g_1}_{{\bm k}|{\bm k}'}(\alpha,\beta|q)- 16\sum_{\nu} \widetilde X^{\g_1\g_2\nu\g_2|\g_3\nu\g_3\g_1}_{k_1k_2k_1k_2|k_3k_1k_3k_1}(\alpha,\beta|q),\nn
\widetilde X^{{\bm\g}|{\bm\alpha}}_{{\bm k}|{\bm q}}(\alpha,\beta|q) &=& Y^{\bm\g}_{\alpha\beta}({\bm k}|q)V_{\bm\alpha}({\bm q}) D( E_{\bm\g}({\bm k})) - ({\bm \gamma},{\bm k})\leftrightarrow({\bm \alpha},{\bm q})\,.
\eea

When implementing the QBE for finite $L$, the parameter $\eta$ in
\fr{AD:Reg} must be kept finite (see e.g.~\cite{SpohnPRE12}). The
results presented in this paper are for $\eta=0.0005$.


\begin{thebibliography}{99}
\bibitem{note:trace}
Only states with a fixed particle number are considered in the trace.
\bibitem{note:imag}
In general the decay times for the real and imaginary parts of the
Green's function are different.
\end{thebibliography}
\end{document}